# A NOTE ON THE PHYSICAL MEANING OF GRADIENT CONSTITUTIVE RELATIONS

A.A. Konstantinidis

Laboratory of Mechanics and Materials, Civil Engineering Department, Faculty of Engineering, Aristotle University of Thessaloniki, GR-54124 Thessaloniki, Greece, akonsta@civil.auth.gr

**Abstract**

Gradient plasticity theory proposed initially by Aifantis and co-workers has proven very useful in problems dealing with material heterogeneity and material instabilities. Although it has been used successfully in many applications by many authors, still some open questions remain unanswered. The modest goal of the present short note is to provide some thoughts on these open questions, providing a new interpretation of the gradient terms leading to a successful interpretation of both extrinsic and intrinsic size effect phenomena.

## 1. INTRODUCTION

The gradient theory of plasticity was proposed by Aifantis [1,2] in the 1980's, motivated by the Van der Waals thermodynamic theory of liquid-vapor transitions and its mechanical counterpart advanced by Aifantis and Serrin [3,4]. By allowing the flow stress to depend on the gradients (up to the second degree and order) of accumulated plastic strain, gradient plasticity was able to describe deformation patterning and the occurrence of shear bands in plastic solids. There were also other models of strain gradient plasticity proposed (for a review see [5] and references contained therein) to consider size effect problems at the micron scale. The simple gradient plasticity model proposed by Aifantis [1,2] was able to dispense with the mesh-size dependence of finite element calculations in the material softening regime, predict the thickness and spacing of shear bands, as well as account for size effects. While this theory was shown to be able to describe many observations at macro and micro scales, when it was proposed it received some criticism mostly on the physical meaning of the gradient terms introduced as well as on its thermodynamic consistency. The latter issue was recently addressed by Gurtin and Anand [6] on the basis of a dissipation inequality and a microforce balance equation earlier introduced by Gurtin [7,8].

The modest goal of the present work is to discuss the physical meaning of the gradient terms introduced in the constitutive relations of gradient plasticity. After a short description of the gradient plasticity formulation, the form and physical meaning of the gradient terms is discussed and a new interpretation is provided in Section 2. With the proposed interpretation extrinsic and intrinsic size effects previously modeled by the gradient plasticity formulation are revisited in Section 3. A comparison with earlier works on scale-dependent constitutive relations is given in Section 4.

## 2. FORMULATION OF THE GRADIENT MODEL

Material heterogeneity consisting of the existence of inclusions, voids, defects, etc, manifests itself, e.g. in simple 1-D tests, through variations in the spatial distribution of strain. The equilibrium equation

$$\partial\sigma/\partial x = 0 \Rightarrow \sigma = \sigma_0, \tag{1}$$

commonly used in such problems implies that a constant stress, i.e. the external stress $\sigma_0$, is applied to every material point. This means that in the cases where the yield stress has not been exceeded, the possible spatial variations of strain are due to material heterogeneity, while if the yield stress value has been exceeded, they could be also due to localized plastic deformation. In both cases one could assume that the material is no longer homogeneous, considering plastic deformation a form of phase change.

For ideal problems where material heterogeneity is not present, a simple stress-strain relation of the form $\sigma = \kappa(\varepsilon)$, where the applied stress affects each material point producing the same amount of local deformation (local strain) would suffice to model the material behavior. In order to study problems in which material heterogeneity is present, non-local models have been proposed. Aifantis [1,2] proposed the introduction in the constitutive relation of the first and the second strain gradient, which in 1-D takes the form

$$\sigma = \kappa(\varepsilon) - c_1|\varepsilon_x| - c_2\varepsilon_{xx} \tag{2}$$

where $\kappa(\varepsilon)$ is the "homogeneous" stress, while the second and third terms of the rhs of Eq. (2) are the gradient terms, with $c_1, c_2$ being the so-called gradient coefficients. Combining Eqs. (1) and (2) the stress-strain constitutive relation takes the form

$$\sigma_0 = \kappa(\varepsilon) - c_1|\varepsilon_x| - c_2\varepsilon_{xx}. \tag{3}$$

In the following we will comment on the purpose of the introduction of the gradient terms and their sign, as well as on the physical meaning of the gradient coefficients $c_1, c_2$.

## 2.1 Sign and form of the gradient terms

First of all, if one wants to model the spatial heterogeneity, he has to introduce in the constitutive stress-strain relation appropriate terms taking into account differences in the material's behavior. These terms should provide a local stress value at each material point enabling either to quantify and model the differences in the material response (spatial distribution of strain) or to "stabilize" the material's response smearing out the local strain differences. The two aforementioned cases can be realized through the use of Eq. (3) by changing the sign in front of the gradient terms. More specifically, a "-" sign in front of the second strain gradient has been used in order to model material heterogeneity, while a "+" sign in order to stabilize the material response. Thus, if one considers the form of Eq. (3) as is, he actually tries to "follow" or model the applied stress, while if this equation is written as $\sigma_0 - c_1|\varepsilon_x| - c_2\varepsilon_{xx} = \kappa(\varepsilon)$, i.e. with a "+" sign in front of the gradient terms, it can be thought of as a an equation modeling the "homogeneous" response $\kappa(\varepsilon)$, i.e. trying to smear out the differences in local strain values. In the following we will consider the constitutive equation as shown in Eq. (3), but the discussion which will follow applies, of course, in both cases.

The introduction of the first and the second gradient of strain in the constitutive relation is crucial since it provides information on the existence of strain differences, provides a quantification through the first gradient, as well as information on whether there is a local minimum (positive value) or maximum (negative value) in the strain distribution through the second gradient. This information is particularly useful when one needs to model the exact material behavior.

## 2.2 Physical meaning of the gradient coefficients

In past works [1-2, 5] the so-called gradient coefficients and especially the coefficient $c_2$ of the second gradient of strain, were related to an "internal" length determined, in general, as $\ell = \sqrt{c_2/S}$, where S is a stress-type quantity, e.g. the Young's modulus, the yield stress, the hardening modulus, etc. First we will deal with the possible physical meaning of the second gradient coefficient.

Actually, as it can be seen from Eq. (3) the gradient coefficient $c_2$ has units of force (N) and this is how it should be treated. It is only indirectly related to a length, as it will be discussed in the following, since this "internal" length is inserted into the constitutive relation through the second spatial derivative of strain $\varepsilon_{xx}$, rather than the gradient coefficient $c_2$.

The second spatial derivative of strain in 1-D actually has the form

$$\varepsilon_{xx} = \lim_{\ell \to 0} \frac{\varepsilon(x-\ell) - 2\varepsilon(x) + \varepsilon(x+\ell)}{\ell^2}. \tag{4}$$

As it can be seen from Eq. (4), it is calculated from the differences in strain between the point into consideration as well as its "adjacent" points being at a distance $\ell$ away from it in the x direction.

In the above relation a crucial point is the fact that this distance $\ell$ tends to zero. This tendency to zero actually depends on the scale of observation of the problem at hand. It is dictated either by the resolution available or the resolution of interest. More specifically, it is related with the detail level with which the microstructure is examined and the resolution at which the strain can be measured. In a macroscopic formulation this distance could be of the order of millimeters, in a mesoscopic scale it could be of the order of nanometers, while in a nanoscopic formulation it could be, if possible, of the order of angstroms. Thus, this distance actually defines the shortest distance between two interfaces or material points, or the size of the vertices of a cubic representative volume element in 3-D that can be used to define a material element. This representative volume is thus a characteristic part of the material dictated by the resolution (scale) in which the constitutive quantities (stress, strain, etc) are measured, i.e. the part of the material for which information on the constitutive quantities can be available.

There is a lower limit for the value of the distance $\ell$ in a continuum formulation and this is the distance between the atoms of the respective material in the crystal lattice, i.e. the interatomic distance. This assumption is even more logical if one considers $c_2$ is the force needed in order for the bond between two atoms to break, in other words the force that leads to plastic deformation. Thus, in the lowest possible resolution, the gradient coefficient $c_2$ will be considered in the next section equal to the force between the atoms of the material which acts as a cohesive force, i.e. the interatomic force.

Following the finite differences scheme, an approximation of the limit in Eq. (4) in a continuum formulation is taken through Taylor series expansion of the strain values in the nominator, i.e.

$$\varepsilon(x-\ell) = \varepsilon(x) - \ell\frac{\partial \varepsilon}{\partial x} + \frac{1}{2}\ell^2 \frac{\partial^2 \varepsilon}{\partial x^2} - \ldots$$
$$\varepsilon(x+\ell) = \varepsilon(x) + \ell\frac{\partial \varepsilon}{\partial x} + \frac{1}{2}\ell^2 \frac{\partial^2 \varepsilon}{\partial x^2} + \ldots$$
(5)

one could easily arrive to the following approximation

$$\frac{\partial^2 \varepsilon}{\partial x^2} \approx \frac{\varepsilon(x-\ell) + \varepsilon(x+\ell) - 2\varepsilon(x)}{\ell^2} \equiv \varepsilon_{\ell\ell} \qquad (6)$$

which will be used as an approximation of the second spatial derivative of strain present in Eq. (3) in the following sections.

### 2.3 Gradient constitutive relations

When the resolution of the problem at hand is not at the atomic scale, but in the meso or macroscale, then the physical meaning of the gradient coefficient $c_2$ and the distance $\ell$ remain the same. But in this case in order for the constitutive relation given in Eq. (3) to be used one has to take into account the specific resolution at hand. In this case the denominator of Eq. (6) is no longer the interatomic distance $\ell$, but a length dictated from the spatial resolution L. From the way the second spatial derivative of strain is calculated at resolution L, the distance $\ell$ present in the denominator of Eq. (6) can be replaced by a scale parameter s denoting the ratio between the resolution L of the problem and the interatomic distance $\ell$, i.e. $s = L/\ell \Rightarrow \ell = L/s$. The same needs to be done also for the first gradient of the strain, thus providing a constitutive relation of the form

$$\sigma_0 = \kappa(\varepsilon) - c_1|\varepsilon_x| - c_2\varepsilon_{xx} = \kappa(\varepsilon) - c_1 s|\bar{\varepsilon}_L| - c_2 s^2\bar{\varepsilon}_{LL} = \kappa(\varepsilon) - c_1\frac{L}{\ell}|\bar{\varepsilon}_L| - c_2\left(\frac{L}{\ell}\right)^2 \bar{\varepsilon}_{LL}, \qquad (7)$$

where $\bar{\varepsilon}_L, \bar{\varepsilon}_{LL}$ are the first and second the spatial derivative of strain with the distance between the material points being equal to L. It should be noted at this point that the resolution L is the distance between material points that can be used in simulations for discretization of the problem at hand.

Returning to Eq. (7), since the gradient coefficient $c_2$ is thought to be the interatomic force, then by considering a simple spring model it would have the form $c_2 = kb = (E\ell)b$, where k denotes the interatomic force constant and b is the lattice distortion, while E denotes the Young's modulus. In order to model plastic deformation that destroys the lattice or separates the material atoms apart by two interatomic distances, i.e. $b \cong \ell$, one needs to set $c_2 = E\ell^2$. In addition, we also assume that the coefficient of the first gradient is equal to the interatomic force constant. Then, Eq. (7) takes its final form

$$\sigma_0 = \kappa(\varepsilon) - c_1'|\bar{\varepsilon}_L| - c_2'\bar{\varepsilon}_{LL}; \quad c_1' = EL, \ c_2' = EL^2. \qquad (8)$$

As mentioned above, earlier the gradient coefficients were thought to be indirectly connected with "characteristic lengths" through relations of the form $\ell_{ch} = c_1/E, \ell_{ch} = \sqrt{c_2/E}$ for the first and second gradient, respectively.

As we have shown in the above discussion a characteristic length is entering the formulation of the model but not through the gradient coefficient. It enters through the gradients of strain and is actually the size of the representative volume element in the specific resolution L at hand; it is characteristic of the resolution of the problem or of the size of material heterogeneity dominating the specific scale of observation. If the scale of observation changes, then the value of this characteristic length should also be changed accordingly. This can easily be understood if one needs to model the same experiment in two different spatial resolutions (scales). The high resolution data contain more complete information than the data obtained using a lower spatial resolution. This difference in the available information leads to the use of a different characteristic length in the constitutive relation, i.e. a different value of the spatial resolution L. The constitutive relation given in Eq. (8) may also provide a means for modeling problems using the same relation from the atomistic scale (L=$\ell$) up to the macroscopic one, thus being useful for the problem of bridging of length scales.

In the next section we provide some examples of using Eq. (8) for modeling size effect problems using the understanding about the gradient coefficients that is proposed in the present work, in order to test its applicability.

## 3. APPLICATION ON EXTRINSIC & INTRINSIC SIZE EFFECTS

### 3.1 Extrinsic Size Effects

The gradient plasticity model given by Eq. (3) was used earlier [9] in order to interpret size effects observed by Morrison [10] and Richards [11] for yield initiation in torsion and bending of mild steel. In this section we compare the results obtained with the ones coming from the proposed formulation. Since the problems at hand consider size effects, the resolution L needed is taken equal in both cases to the smaller difference in the characteristic dimension of the specimens used. This is due to the fact that the available data do not provide any information on the resolution with which the respective yield stress measurements were taken.

*3.1.1 Yield initiation in torsion*

Morrison [10] performed a series of careful torsion tests on the yield behavior of plain carbon steel cylindrical specimens of different size. These results were modeled by Tsagrakis et al [9] using Eq. (3), which was used transformed into the constitutive relation between shear stress and shear strain through the use of a yield criterion very close to the Tresca yield criterion. Then a strength of materials approach led to the yield stress Y of a specimen with radius α given by the relation

$$\frac{Y(\alpha)}{\sigma_0} = \Lambda \left( \frac{\alpha^2}{\alpha^2 + \bar{c}_2/G + \alpha\,\bar{c}_1/G} \right), \qquad (9)$$

with G denoting the shear modulus and $\sigma_0$ the tensile yield stress. The values of the gradient coefficients and the respective internal lengths $\bar{c}_1/G = -0.38$ mm and $\sqrt{c_2/G} = 0.476$ mm, as well as the coefficient $\Lambda = 0.516$ introduced due to the yield criterion, were calculated by fitting the experimental data (blue curve in Fig. 1).

As mentioned above, the resolution L provided by the available experimental data is the smaller difference in the characteristic dimension of the specimens used. This macroscopic size in this case is the radius $\alpha$ of the specimens and, thus, the resolution L is considered to be $L = 0.98$ mm. The values of E and G needed were taken from the literature as $E = 210$ GPa and $G = 74$ GPa, providing for the gradient coefficients the values $c_1 = -205.8$ kN/mm and $c_2 = 201.68$ kN. It is noted at this point that, as also the case with the work of Tsagrakis et al [9], the values of the gradient coefficients were multiplied with the factor $\Lambda^2$ of the yield criterion, which in this case was the Tresca yield criterion with $\Lambda = 0.5$. The predictions of the formulation are shown with the red curve in Fig. 1.

From the comparison between the predictions of the two formulations it is clear that they are equally able to predict the size effect present in Morrison's experimental data. But it should be noted that with using the model proposed in this work the experimental data can be modelled sufficiently well while the values of the gradient coefficients are not calculated through any kind of fitting. Only the value of the moduli and the observation of the respective resolution are needed for calculating their values.

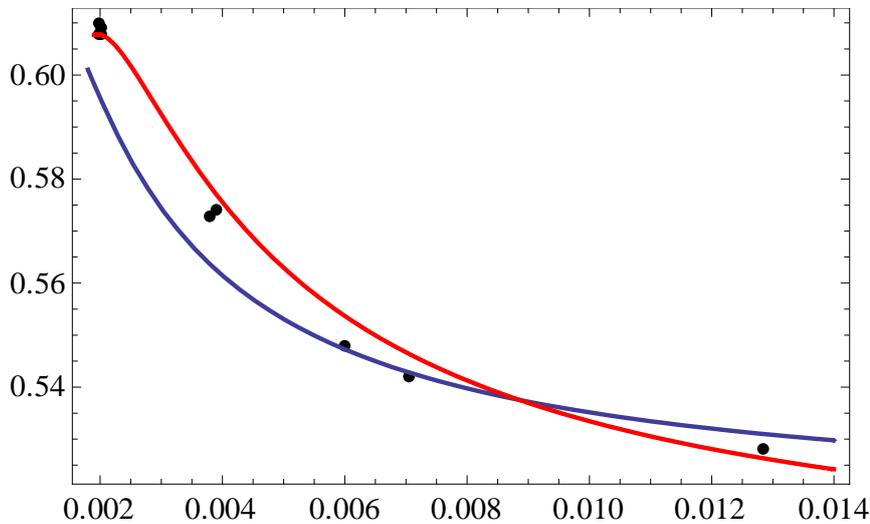

**Figure 1:** Comparison between the predictions of gradient theory (blue curve) and the formulation proposed in this work (red curve) on size effect data on yield stress in torsion experiments of plain carbon steel [10].

*3.1.2 Yield initiation in bending*

Richards [11] performed a series of pure bending tests for geometrically similar mild steel beam specimens of different size. The bending data were modeled again by Tsagrakis et al [9] using Eq. (3). The dependence of the yield stress Y on the specimen depth h through a strength of materials approach was found to be given by

$$Y(h) = \sigma_0 \left( \frac{h/2}{h/2 + c_1/E} \right), \tag{10}$$

with $\sigma_0$ denoting the tensile stress. The values $c_1/E = -1.125$ mm of the internal length, as well as the tensile stress ($\sigma_0 = 225.6$ MPa), were calculated by fitting the experimental data (blue curve in Fig. 2).

Within the proposed formulation, the value of the gradient coefficient is calculated by the mild steel Young's modulus ($E = 210$ GPa) as well as the value of the problem's resolution L which in this case was the smallest difference between the values of $h/2$ (the characteristic dimension of the specimens), i.e. $L = 1.25$ mm. Using these values the gradient coefficient in this case is $c_1 = -262.5$ kN/mm, and the predictions of the model are shown with the red curve in Fig. 2.

Again, the predictions are quite satisfactory, although the value of the gradient coefficient did not come from fitting the data.

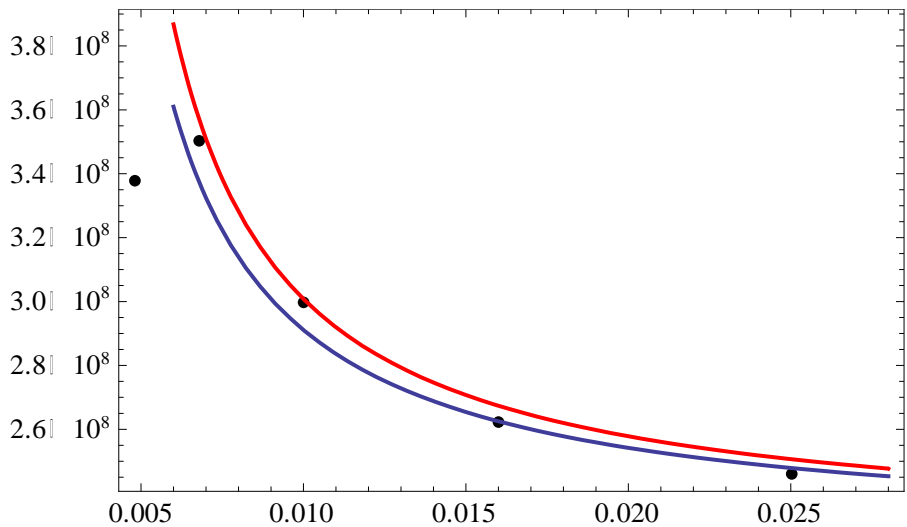

**Figure 2:** Comparison between the predictions of gradient theory (blue curve) and the formulation proposed in this work (red curve) on size effect data on yield stress in bending experiments of mild steel [11].

## 3.2 Intrinsic Size Effects

In contrast to the extrinsic size effects (dependence on specimen size) studied in the previous subsection, the proposed formulation can be applied in intrinsic size effects as well (dependence on specimen microstructure size). The most studied size effect of this category is the well known Hall-Petch behavior [12-13], i.e. the dependence of the hardness or yield strength on grain size, as well as the so-called "inverse" Hall-Petch behavior. The latter has been studied in the past [14,15] using as a starting point the theory of gradient plasticity modified through the substitution of the gradient term with a scalar scale-dependent one [16-17]. In this subsection the formulation proposed in this article is used, providing a mechanics-based framework for dealing with the aforementioned topics.

The gradient plasticity expression of Eq. (3) is used, i.e.

$$\sigma_0 = \kappa(\varepsilon) - c_1 \frac{L}{\ell} |\bar{\varepsilon}_L| - c_2 \left(\frac{L}{\ell}\right)^2 \bar{\varepsilon}_{LL}. \tag{11}$$

In this case the resolution L is not identified with a macroscopic specimen dimension, but with the grain size d, leading to an equation of the form

$$\sigma_0 = \kappa(\varepsilon) - c_1 \frac{d}{\ell} |\bar{\varepsilon}_d| - c_2 \left(\frac{d}{\ell}\right)^2 \bar{\varepsilon}_{dd} = \kappa(\varepsilon) - c_1 \frac{d}{\ell} \left|\frac{\partial \bar{\varepsilon}}{\partial d}\right| - c_2 \left(\frac{d}{\ell}\right)^2 \frac{\partial^2 \bar{\varepsilon}}{\partial d^2}. \tag{12}$$

Equation (12) holds for various values of the external stress $\sigma_0$ and the corresponding values of the homogeneous stress $\kappa(\varepsilon)$. Thus, it also holds for the value of the yield stress $\sigma_y$, which would be constant and equal to the homogeneous yield stress $\kappa_y$ in the absence of material heterogeneity. In this case, and taking into account that again $c_1 = E\ell$ and $c_2 = E\ell^2$, Eq. (12) leads to an expression of the form

$$\sigma_y = \kappa_y - Ed \left|\frac{\partial \varepsilon_y}{\partial d}\right| - Ed^2 \frac{\partial^2 \varepsilon_y}{\partial d^2}, \tag{13}$$

which, by assuming that the yield stress is the stress value where the stress vs. strain graph departs from linearity (similar to the case where the yield stress is defined as the stress for leading to 0.2% plastic deformation), i.e. $\sigma_y = E\varepsilon_y$, gives

$$\sigma_y = \kappa_y - d \left|\frac{\partial \sigma_y}{\partial d}\right| - d^2 \frac{\partial^2 \sigma_y}{\partial d^2}. \tag{14}$$

Assuming now that the yield stress of the homogeneous case is constant (not depending on grain size), Eq. (14) has the following solution

$$\sigma_y = \kappa_y + \lambda_1 \cos(\ln d) + \lambda_2 \sin(\ln d), \tag{15}$$

with $\lambda_1, \lambda_2$ phenomenological constants that need to be determined by comparison of the predictions of Eq. (15) with experimental data.

In the case of nanocrystalline Cu [18], the fitting provided the values $\kappa_y = 218$ MPa, $\lambda_1 = -34$ MPa and $\lambda_2 = -145$ MPa and is shown in Fig. 3, in good agreement with experimental measurements. It is noted at this point that the dependence of yield stress on grain size comes from a mechanics-based formulation, i.e. gradient plasticity, in contrast with other models that a priori use the semi-empirical Hall-Petch rule of yield stress dependence on the inverse square root of the grain size.

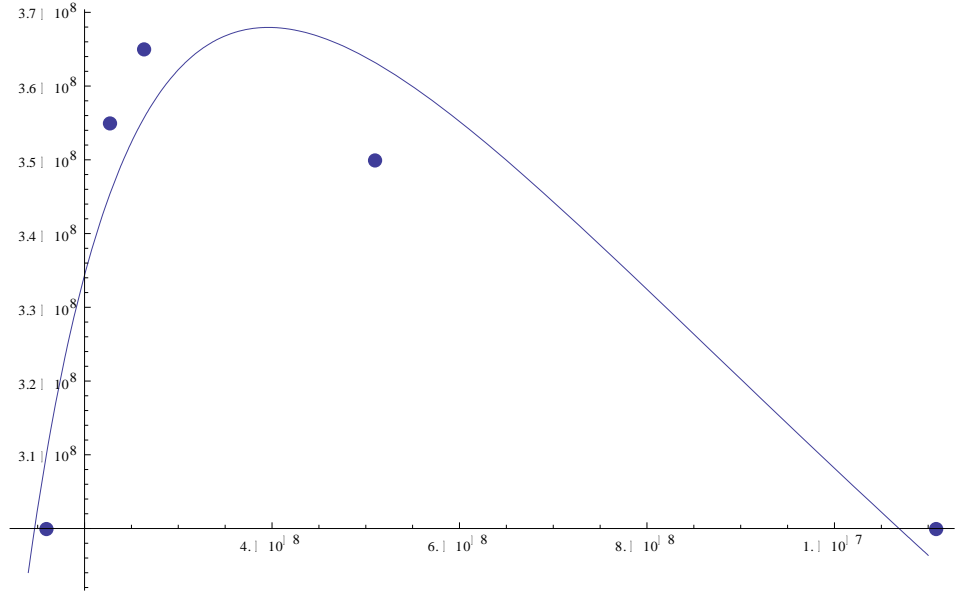

**Figure 3:** Predictions of the proposed formulation for the yield stress dependence in nanocrystalline Cu [18].

## 4. COMPARISON WITH SCALE-DEPENDENT CONSTITUTIVE RELATIONS

In previous works by the author and co-workers [16-17, 19-22] wavelet analysis was used in order for scale-dependent constitutive relations to be derived from the gradient plasticity constitutive equation given by Eq. (3), by replacing the gradient term with a scalar one containing a scale parameter s. The scale-dependent constitutive relation [16-17,19-22] derived from a simplified form of Eq. (3), with $c_1 = 0$, i.e.

$$\sigma_0 = \kappa(\varepsilon) - c\varepsilon_{xx} = \kappa(\varepsilon) - c\left(\frac{L}{\ell}\right)^2 \bar{\varepsilon}_{LL}, \qquad (16)$$

had the form

$$\sigma_0 = \kappa(\varepsilon) + c\frac{\varepsilon}{2s'^2}\left[1 + 2\log\left(\frac{2\varepsilon s'\sqrt{\pi}}{s_0}\right)\right], \qquad (17)$$

where $s'$ is a scale factor and $s_0$ the total displacement. Equation (17) was derived by assuming that the localized strain distribution was approximated by the wavelet representation $\delta_s$ of the $\delta$-function of the form

$$\delta_s(x) = \varepsilon(x) = \frac{s_0}{2s\sqrt{\pi}}e^{-x^2/4s^2}. \qquad (18)$$

The gradient terms of Eqs. (16) and (17) are compared in order for the scale factor $s'$ of Eq. (17) to be connected with the factor $L/\ell$ present in Eq. (16) as

$$-c\left(\frac{L}{\ell}\right)^2 \bar{\varepsilon}_{LL} = -c\left(\frac{L}{\ell}\right)^2 \frac{\bar{\varepsilon}(x+L)+\bar{\varepsilon}(x-L)-2\bar{\varepsilon}(x)}{L^2} \equiv c\frac{\varepsilon}{2s'^2}\left[1+2\log\left(\frac{2\varepsilon s'\sqrt{\pi}}{s_0}\right)\right]. \qquad (19)$$

In the above equation it is assumed that $\bar{\varepsilon}(x)$ has the form given in Eq. (18). Then after some straightforward mathematical manipulation, Eq. (19) gives

$$-\frac{e^{-\frac{(L-x)^2}{4s'^2}}+e^{-\frac{(L+x)^2}{4s'^2}}-2e^{-\frac{x^2}{4s'^2}}}{L^2} \equiv \frac{\left(1-x^2/2s'^2\right)e^{-\frac{x^2}{4s'^2}}}{2s'^2}. \qquad (20)$$

Evaluating the two gradient terms of Eq. (20) at the center of the localization, i.e. for $c_1 = 0$, Eq. (20) gives

$$\frac{L^2}{4s'^2} \equiv 1 - e^{-\frac{L^2}{4s'^2}} \qquad (21)$$

which always holds true for $s' > 2L$. This result means that the formulation of scale-dependent constitutive equations used earlier [16-17, 19-22] for modeling size effect problems is exactly the same with the formulation proposed herein when the scale parameter $s'$ is take to be greater than twice the resolution length L.

## 5. CONCLUSIONS

We have discussed the physical meaning of the gradient coefficient c entering the constitutive relations of gradient plasticity proposed by Aifantis [1,2]. Its physical meaning is the interatomic forces of the material at hand. In addition, a distinction between modeling the applied stress, or the true material behavior, and modeling the homogeneous response, or the ideal material behaviour, is made and this distinction leads to the conclusion that the sign of the gradient coefficient is always positive, but the sign of the gradient term as a whole changes, depending on what we try to model. Another point of interest is that the characteristic length related with the gradient coefficient is actually dictated by the scale of observation (spatial resolution) and enters the formulation due to the way the second spatial derivative of strain is calculated, rather than the gradient coefficient itself. The proposed formulation is not contradictory to previous works on gradient plasticity theory, but is rather more definite, with the physical meaning of the gradient coefficient well defined, and the possibility for multiscale modeling (or attacking the problem of bridging of length scales) provided through the use of the spatial resolution L as the characteristic length in the constitutive equation given in Eq. (6).

A more rigorous treatment on the issues discussed herein is obviously needed, but this is beyond the scope of this short note which aims at providing a few thoughts answering the main questions initially raised pertaining to the theory of gradient plasticity proposed by Aifantis [1,2], that may be useful to the researchers that are now entering this field.